\documentclass{pnastwo}
\usepackage{graphicx,amsmath}
\usepackage{graphicx}
\usepackage{dcolumn}
\usepackage{bm}
\usepackage{url}
\usepackage{color}
\usepackage{floatflt}
\usepackage{hyperref}
\usepackage{amssymb,amsmath}
\usepackage{graphicx}
\usepackage{verbatim}
\usepackage{graphics}
\usepackage{wrapfig}
\usepackage{subfigure}
\usepackage{float}

\begin{document}


\title{Atmospheric stability shapes strategies for fungal spore liberation}
\title{Turbulence dictates survival of fungal spores in the atmosphere}
\title{Timing of fungal spore release dictates survival during atmospheric transport}

\author{Daniele Lagomarsino Oneto\affil{1}{CNRS and Universit\'e C\^ote d'Azur, Institut de Physique de Nice UMR7010, parc Valrose 06108 Nice, France}, Jacob Golan\affil{2}{Departments of Botany and Bacteriology, University of Wisconsin-Madison, Madison, WI, USA}, Andrea Mazzino \affil{3}{Department of Civil, Chemical, and Environmental Engineering, University of Genova and INFN, via Montallegro 1, 16145 Genova, Italy}, Anne Pringle \affil{2}{Departments of Botany and Bacteriology, University of Wisconsin-Madison, Madison, WI, USA},  Agnese Seminara \thanks{To whom correspondence should be addressed. E-mail:agnese.seminara@unice.fr}\affil{1}{CNRS and Universit\'e C\^ote d'Azur, Institut de Physique de Nice UMR7010, parc Valrose 06108 Nice, France}}

\maketitle

\begin{article}
\begin{abstract}
The fungi disperse spores to move across landscapes and spore liberation takes different patterns. While many species release spores intermittently, others release spores at specific times of day or night according to intrinsic rhythms. Despite intriguing evidence of diurnal rhythms, why the timing of spore liberation would matter to a fungus remains an open question. Here we use state-of-the-art numerical simulations of atmospheric transport with meteorological data to follow the trajectory of many spores released in the open atmosphere at different times of day, during different seasons and at different locations across North America. While individual spores follow unpredictable trajectories due to turbulence, in the aggregate patterns emerge: statistically, spores released during the day fly for several days, while spores released at night return to the ground within a few hours. 
Differences are caused by intense turbulence during the day and weak turbulence at night. 
The pattern is widespread but its reliability varies, for example, day/night patterns are stronger in southern regions, where temperatures are warmer. Results provide a set of testable hypotheses explaining intermittent and regular patterns of spore release as strategies to maximize spore survival in the air.
Species with short lived spores reproducing where there is strong and regular turbulence during the day, for example in Mexico, will maximize survival by routinely releasing spores at night. Where cycles are weak, for example in Canada during spring, there will be no benefit to releasing spores at the same time every day. 
We also challenge the perception of atmospheric dispersal as risky, wasteful, and beyond control of a sporocarp; our data suggest the timing of spore liberation may be finely tuned by a fungus to maximize fitness during atmospheric transport.
\end{abstract}

A careful reading of the natural history of fungal spore liberation reveals nontrivial patterns. Spores may be released at specific times of the day according to an internal clock, or may be released according to fluctuations in the environment. Some species display regular, nearly circadian rhythms, e.g.~the asexual spores of the powdery mildew Genus species are mostly released at midday  (\cite{mildew} and references therein). Other species release spores preferentially at night or early morning, including the plant pathogens  \emph{Mycosphaerella fijiensis} (causing wheat leaf blotch and black leaf streak in bananas) \cite{Meredith_Lawrence1973,myco_book,regular5,meredith2009}, \emph{Giberella zeae} (causing crow rot in cereals) \cite{Paulitz1996,Gilbert2008}, \emph{Venturia inequalis} (the apple scab) \cite{venturia1,regular2} as well as several tropical species \cite{nocturnal1,nocturnal2}. Or patterns may be more complex, e.g. the causal agent of blackleg, \emph{Leptosphaeria maculans}, 
seems to follow different diurnal rhythms in different regions and seasons, with most spores liberated in the morning in England \cite{huang2005} or at night in Canada \cite{guo2005}, or early afternoon in Western Australia \cite{savage2012}. 
Alternatively, there may be no regular pattern, e.g.~in different studies of \emph{L maculans} in Western Australia spores were released intermittently, regardless of location or month \cite{khangura2007,mcgee1977}. \\
Many authors have attempted to connect the diversity of spore liberation patterns to specific environmental conditions by drawing correlations with local temperature, humidity and wind speed (reviewed in \cite{mildew,review_effects_meteo,aylor_annurev}). 
For example, asexual spores of \emph{Helminthosporium mayds} (syn. = Bipolaris maydis) and \emph{Alternaria spp.}~are detached from the substrate by intense wind gusts
(e.g.~\cite{aylor1974,rotem}).
Despite a wealth of data describing different diurnal or nocturnal patterns of spore liberation, and a nascent understanding of their importance for spore survival \cite{nathan2011,savage,savage2010} the causes of these patterns remain obscure \cite{savage2012,review_effects_meteo}. \\

Atmospheric dispersal is assumed to be both common and dangerous. Cellular material makes up about 25\% of the atmospheric particulate and 3\% to 11\% by weight \cite{boreson,jaenicke}, with crucial implications for health, agriculture and climate. But of the estimated $\sim10^{21}$ cells riding the atmosphere annually, only a small fraction may survive the journey. Exposure to UV light damage, and to uncontrolled fluctuations in temperature and humidity are the main threats that limit the lifespan of spores in the open atmosphere \cite{norros_survival}. \\
To date research has focused on the reach of a spore, in other words, biologists have sought to understand how far a spore will move before deposition, and paid attention to distances traveled. For example, it is generally recognized that spores and seeds released when turbulence is intense are more likely to undergo long distance dispersal \cite{savage,aylor,dandelion}. 
More rarely considered is the lifespan of a fungal spore during transport in the open atmosphere. But in fact spores in flight are exposed to e.g.~UV light damage, and may survive for a time that ranges from less than an hour \cite{norros_survival} to several weeks \cite{Mundt2009}, and possibly even longer \cite{survival_stratosphere}. Spore lifetime directly affects fitness: a spore that dies in the atmosphere will have zero fitness, even if it ultimately settles back to the ground. \\

The duration of a spore's journey in the atmosphere dictates its chances of survival: spores survive in the atmosphere if they return to the ground during their lifetime. 
But the duration of the journey for individual spores is inherently unpredictable due to turbulence: two identical spores released from a single sporocarp may take radically different paths \cite{fgv}. 
However, the average flight time for a group of spores released simultaneously from the same location may follow a specific pattern,
which is often studied in the context of aerosol science (and named residence time or flight time, see e.g.~\cite{aerosol_book}). 
The flight time of large aerosols (diameter $5-20\,\mu m$, similar to a typical fungal spore) results from the balance between two opposite forces: gravity causes particles to sediment downward, and turbulence keeps them aloft~\cite{aerosol_book}. Hence for example residence time for larger particles is shorter~\cite{aerosol_book,tellus1974,denjean}. 
To make the above argument quantitative, aerosol science often assumes that the dynamics has reached an equilibrium because e.g.~particles take off from a large area that does not vary in time. 
But this is not the case for fungi, which are discrete entities, distributed in irregular patches and may produce and release spores at one point in time only. 
A recent model~\cite{grisha} considers particles released at one time in the idealized case of a vertically infinite neutral atmosphere, i.e.~where the intensity of turbulence increases linearly with altitude, and does not change in time. A main conclusion of their idealized model is that the flight time becomes infinitely long when turbulence is stronger than sedimentation.  
More realistic analyses of spore flight time considering variations with season, geography, and state of the atmosphere, require massive numerical simulations using meteorological data. \\

By combining state-of-the-art numerical simulations of atmospheric particle transport with simplified models of atmospheric turbulence, and by explicitly considering spore lifespan, we discover the timing of spore liberation dramatically influences the effective reach of viable spores. Manipulating the timing of spore liberation will dramatically influence fitness. We find (i) the average duration of a spore's flight depends on when it is released; by explicitly defining fitness as the fraction of spores that sediment during their lifetime we discover patterns in flight time cause specular patterns in fitness. (ii) Turbulence dominates vertical transport and thus dictates flight time in realistic conditions, reminiscent of previous results limited to idealized transport models. 
(iii) The cyclical nature of turbulence drives observed patterns in fitness: typically, turbulence is stronger during the day versus at night. The strength and reliability of this diurnal cycle of turbulence varies with geography and season. (iv) When and where the cyclical pattern of turbulence is unreliable, a direct measure of the local intensity of turbulence will be a better guide than time of day to maximise fitness. 
Results provide a set of testable hypotheses to understand observed patterns of spore release:
(1) Releasing spores at specific times of the day is beneficial for species living in regions where the atmosphere cycles regularly. (2) For these species, short-lived spores should be released at night, while long lived spores can be released during the day. (3) Intermittent patterns of spore release may emerge as an adaptation to an environment where the diurnal cycle of turbulence is disrupted. \\
While atmospheric transport is assumed to be unpredictable, and fungi are assumed to have little control over it, spore discharge itself appears finely tuned to maximize individual fitness \cite{marcus_shape,apical_ring,puffing,review_frances,book_chapter,annurev}. Our results demonstrate that fungi may still maximize one aspect of fitness by strategizing timing of spore release.  
 
 \section{Results}
The duration of a spore's flight in the atmosphere can be controlled by the timing of spore release. 
To determine the statistics of spore flight time, we follow the trajectories of many spores released instantaneously from single sites. 
We model an array of 10 locations in North America (see Supplementary Figure 1 and Supplementary table), releasing groups of 100,000 spores from the first layer of the atmosphere closest to the soil every 3 hours over the course of four different months (January, April, July and October 2014), resulting in a total of 9600 numerical simulations. Our simulations track the Lagrangian trajectories of spores in the atmosphere using meteorological data publicly available from NOAA and the software HYSPLIT \cite{narr,hysplit1,hysplit2,hysplit3,hysplit4}, see Materials and Methods and Supplementary information. 
Particles are modeled as tracers carried across the atmosphere both vertically and horizontally, with a specific gravitational settling velocity. 
Spores are carried by the large scale wind field, coming from meteorological models,
and are additionally kicked and buffeted by turbulent fluctuations. Turbulence is modeled as a correlated stochastic process akin to a simple diffusion, with an effective diffusivity (\emph{eddy diffusivity}) that depends on height. 
To mimic deposition, spores remaining or returning to the first layer closest to the soil are randomly removed from the simulation and returned to the ground with a constant rate proportional to the deposition velocity. Our simulations focus on the large scales representing the journey of spores that travel in the open air, and do not resolve the details of release and deposition within the canopy. 
We record the duration of each particle's trajectory in the open air from take off to landing and analyze the statistics of each group of 100,000 spores. 
We consider deposition velocity equal to sedimentation velocity, as suitable for large particles \cite{aerosol_book} and we set it to $6\,mm/s$ (corresponding to an equivalent sphere of radius 6$\mu m$ and density equal to the density of water). We follow each group of spores for 6 weeks; by then, most spores have sedimented to the ground; a small fraction of spores (on average about 16\%) escapes into the stratosphere and is rapidly carried outside of the computational domain by strong geostrophic winds, see supplementary Figure 2. \\
\begin{figure}[h!]
\begin{center}
\includegraphics[width=0.5\textwidth]{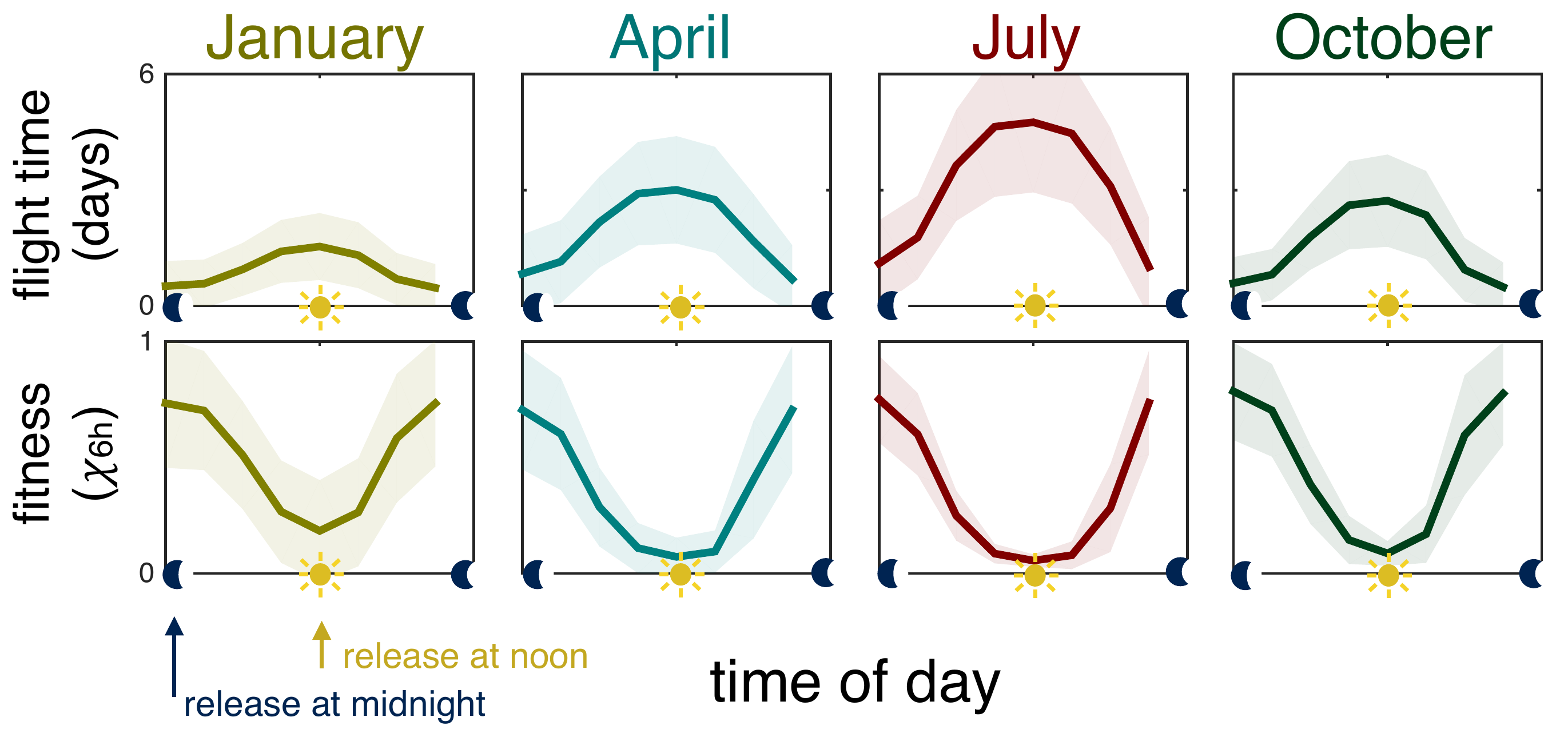}
\caption{Spore survival depends on time of liberation. Top row: Average flight time from liberation to deposition as a function of the time of day. Eight take-offs were simulated each day at ten locations for the entire months of January, April, July and October 2014. Bottom row: Fitness $\chi_{6h}$, defined as the fraction of spores deposited within their lifetime $\tau$: $\chi = \int_0^\tau p(t) dt$ with $\tau=$~6 hours.
}
\end{center}
\end{figure}
We find that spore release at different times of the same day results in dramatic oscillations in flight time ranging from less than an hour to several days (Figure 1, top row), consistent with previous results in the context of aerosol science \cite{aerosol_book}. Longer flight times are observed in the summer (Figure 1, top row), confirming a previously observed seasonal pattern in residence time of abiotic particles \cite{balkanski_jacob}. We define one aspect of \emph{fitness} 
\cite{pringle_and_taylor} as the fraction of particles deposited within their lifetime $\tau$ and indicate it with the symbol $\chi_\tau$ ($w$ is commonly used for fitness in evolutionary biology but here we reserve the symbol $w$ for velocity, as is standard in the physics literature). 
Similar to the results for flight time, fitness also undergoes oscillations (Figure 1, bottom row for $\tau=6$ hours). Note that peaks in flight time correspond to minima in fitness. Indeed, maximum flight times are on the order of several days and in these conditions most of the spores die in flight, since their lifetime is $6$ hours. Conversely at night flight times are minima and spore survival is maximum. Hence short-lived spores should be released at night in order to maximize fitness.
Similar relationships hold for different choices of lifetime $\tau$, but fitness tends to flatten out for longer lived spores, which can survive even very long flights (see Supplementary Figure 3 for results with $\tau=2$~days and $\tau=2$~weeks). Hence timing of spore release only weakly affects survival of long lived spores, which may be released at any time of the day: other aspects of fitness will shape their liberation patterns, for example the need to maximize dispersal range. 
Demographic variables, and specifically the temporal viability of spores, emerge as critical controls on successful dispersal.

\begin{figure}[h!]
\begin{center}
\includegraphics[width=0.5\textwidth]{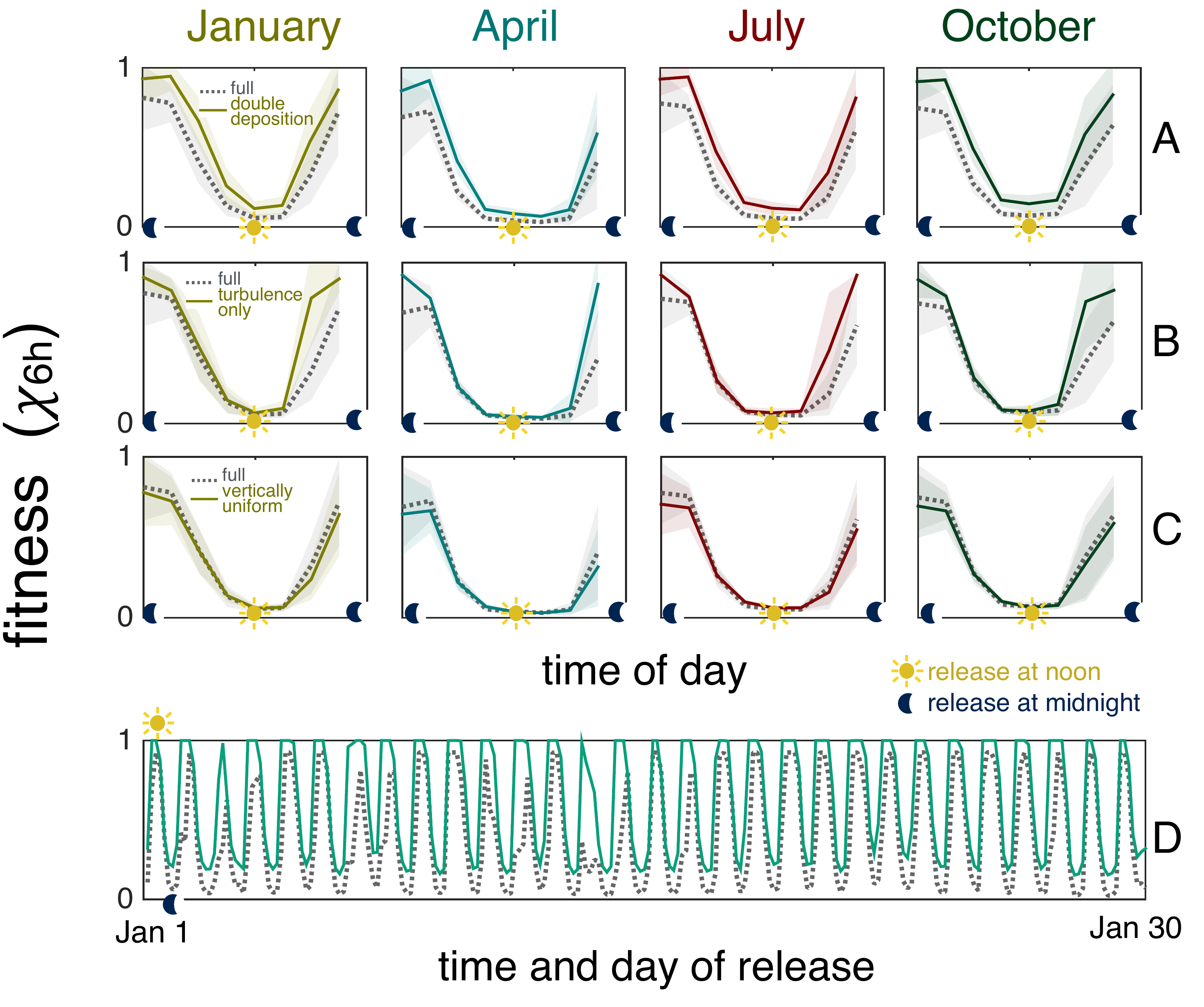}
\caption{
Turbulence dictates fitness: wind, the details used to model turbulence, and deposition efficiency only slightly affect results. In all panels, grey dotted lines represent fitness averaged across all simulations performed in the same month, as a function of the timing of spore release with $\tau=$~6 hours, computed from the full HYSPLIT simulations described in the text. Gray shading represents the standard deviation of fitness over all simulations. Solid lines represent fitness calculated from:  (A) HYSPLIT simulations where the efficiency of deposition is doubled; (B) simplified HYSPLIT simulations without large scale meteorological winds; (C) HYSPLIT simulations with an eddy diffusivity that varies in time but not in z for the entire troposphere; (D) one dimensional finite difference simulations implementing the eddy diffusivity model only (no gravitational settling, no large scale meteorological winds, no horizontal displacement). Note how the model that implements only turbulence reproduces nearly exactly the full model.  
All simulations start from the same location (10) in Mexico (see Supplementary Table 1).  
\label{fig2}}
\end{center}
\end{figure}
Oscillations of flight time and fitness are qualitatively robust to small variations of sedimentation and gravitational settling and they will vary quantitatively. 
Doubling deposition or sedimentation results in shorter flight times and therefore increase in fitness (see Figure 2A and supplementary Figure 4).
This result has been previously recognized in the literature for fungal spores (e.g.~\cite{tellus1974}) and is well known in the context of aerosol science (e.g.~\cite{aerosol_book}). 
A substantial difference occurs for much larger sedimentation of $6\,cm/s$ (corresponding to a sphere with diameter $40\,\mu m$), where sedimentation dominates the dynamics, as well known for seeds \cite{nathan2011} (supplementary Figure 4). 
In the following we will not further discuss this regime, and focus on the more typical case with spores smaller than $\sim 20\,\mu m$ in diameter. Next, we use our simulations to trace these fitness oscillations back to turbulence.

Flight times and fitness depend mainly on turbulence.
Figure 2B shows that fitness is only weakly sensitive to the wind that spores experience along their trajectory because typically, turbulence dominates over vertical wind. Horizontal winds do not affect the vertical dynamics either, because meteorological parameters vary slowly in space. 
To test the robustness of our turbulence model, we implemented two closures for subgrid fluctuations (see Materials and Methods). Both closures result in the same qualitative results (supplementary Figure 5). We next tested whether the exact profile of these phenomenological expressions is relevant for understanding the mechanism. To this end, we further simplified the HYSPLIT  simulation by using a vertically uniform eddy diffusivity resulting from the average over the entire column; the results accord well with the full numerical simulation (Figure 2C). 
These results suggest that the intensity of turbulence dictates fitness almost entirely. To test this idea, we temporarily left the HYSPLIT framework, and modeled the rate of change of spore concentration over the vertical direction, using solely the turbulent model extracted from HYSPLIT (Eulerian eddy diffusivity model, see Materials and Methods).
Fitness is well approximated by this bare bones model, confirming that the intensity of turbulence is  the single major parameter dictating fitness.

\begin{figure}[h!]
\begin{center}
\includegraphics[width=0.5\textwidth]{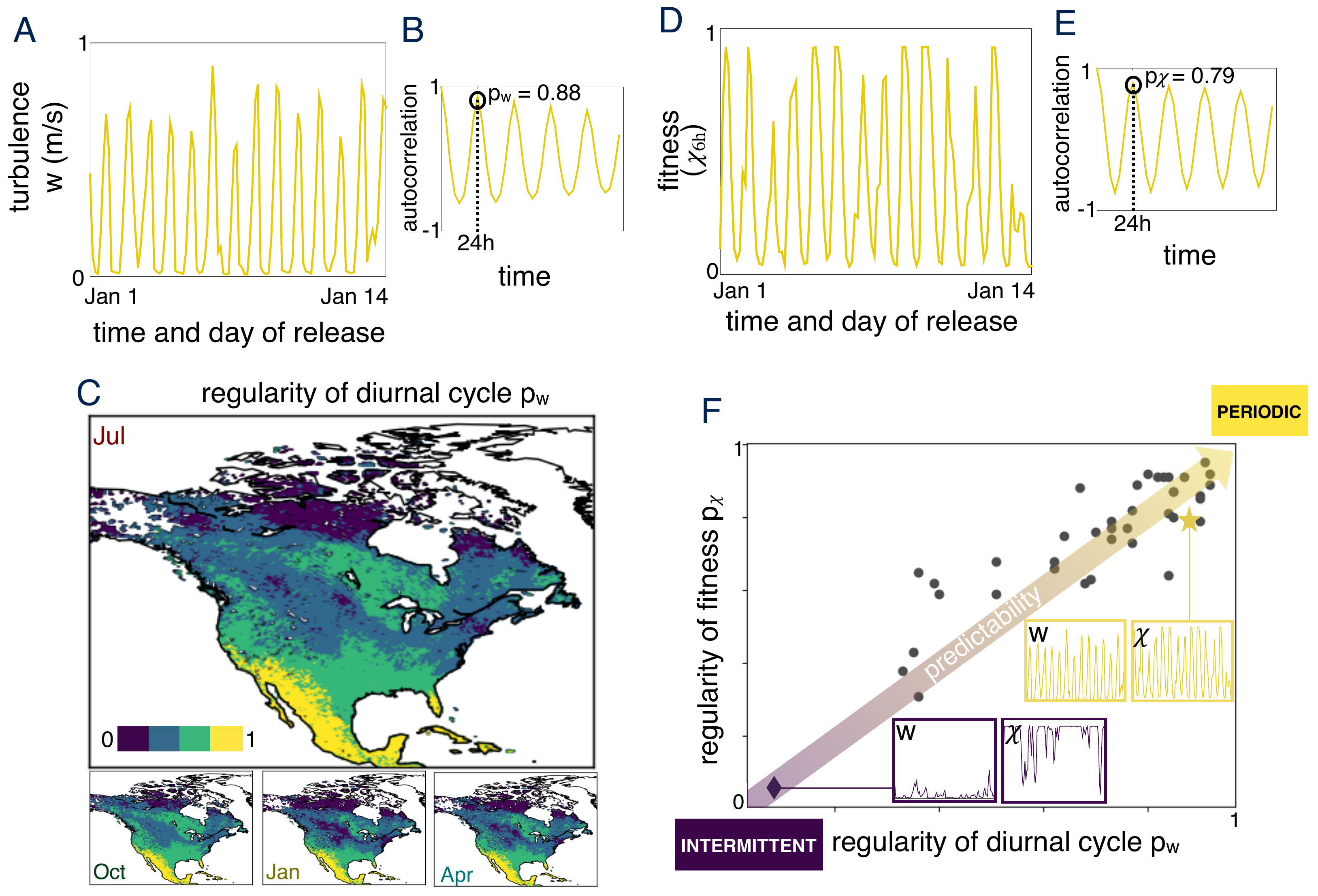}
\caption{ 
Atmospheric turbulence undergoes a diurnal cycle that varies according to geography and season. (A) Magnitude of vertical turbulent velocity (\emph{w}) during two weeks in January in Mexico (in 2014 at location \#10 in Mexico, Supplementary Table 1). 
(B) The corresponding autocorrelation function $<(w(t)-\bar{w})(w(t+t')-\bar{w})/\sigma^2_w>$, where $< . >$ denotes average over 10 days. We define an index $p_w$ (for periodicity) measuring the autocorrelation at 24h, a measure of the reliability of the diurnal cycle. For a perfectly periodic signal $p_w=1$, whereas for an intermittent signal $p_w=0$. 
(C) Map of $p_w$ computed from meteorological datasets, color coded from yellow (regular) to purple (intermittent), for the entire North American continent and different seasons. 
(D) Regular oscillations of fitness $\chi_{6h}$ and (E) its autocorrelation $<(\chi_{6h}(t)-\bar{\chi}_{6h})(\chi_{6h}(t+t')-\bar{\chi}_{6h})/\sigma^2_\chi>$ calculated for the same simulations and as described for turbulence in panel (B). The autocorrelation of $\chi_{6h}$ at 24h defines the index $p_\chi$ for fitness, similar to the index $p_w$ for velocity.
(F) Index $p_\chi$ for fitness at 6~hours  is positively correlated to index $p_w$ for turbulence. The lower left corner corresponds to highly intermittent cases (purple diamond is location \#1 in Canada in January), while the top right corresponds to extremely regular conditions (yellow star is location \#10 in Mexico in January, shown in panels A,D). 
}
\end{center}
\end{figure}
Oscillations in fitness are caused by alternations between strong turbulence during the day and weak turbulence during the night. During the day the sun warms up the atmosphere and soil, the soil warming up faster than the air. 
The soil in turn warms the lowest layers of the atmosphere and because warm air tends to rise, it powers thermal convection and intense turbulence. Releasing spores in these conditions will cause them to be carried in the upper layers of the atmosphere, resulting in long flights during the day.
Conversely, during the night, the lack of sun causes fast cooling of the soil and hence of the lowest layers of air, often causing stable stratification and weak turbulence. Spores released in these stable conditions never reach large altitudes and return to the ground faster. To understand where this diurnal cycle may be strongest, we extracted from the full meteorological dataset the intensity of turbulent fluctuations: we indicate with $w$ the standard deviation of fluctuations in vertical air speed, mediated in height and computed within HYSPLIT.
Larger values of $w$ correspond to greater intensities of turbulent fluctuations. In Figure~3A we show $w$ measured from  location \#10 in Mexico during the first 2 weeks of January. Turbulence displays regular diurnal oscillations over the entire period. To quantify the consistency of this diurnal cycle, we define an index $p_w$ as the autocorrelation of turbulence $w$ at 24~h (Figure 3B). This index ranges from 1 for a perfectly periodic signal to 0 for an irregular or \emph{intermittent} signal. 
We next compute $p_w$ for the entire North American continent for the months of January, April, July and October. We approximate $w$ as the standard deviation of the convective velocity, a parameter that is available from the meteorological dataset, and does not require HYSPLIT. 
Although the diurnal cycle is widespread, its consistency or reliability varies with geography and season (Figure 3C). 
To verify whether the diurnal cycle of turbulence affects fitness, we compute a second index $p_\chi$ analogous to $p_w$ and measuring periodicity of fitness for each of the 10 starting locations and 4 different months we simulated previously. Figure 3D-E shows one example of fitness $\chi$ and its autocorrelation defining $p_\chi$, for the same location/weeks illustrated in Figure 3A-B. Next we correlate $p_w$ with $p_\chi$ and we find that periodicity of turbulence correlates well with periodicity of fitness (Figure 3F). 
This correlation is extremely useful, as the index $p_w$ can be computed from meteorological data, and this index can be used to estimate $p_\chi$ with no need of simulating spore trajectories.  
This analysis suggests that when and where turbulence is periodic $p_w\sim 1$, fitness is also periodic $p_\chi\sim 1$ and in these regions marked in yellow/green in Figure 3C, releasing spores at specific times of the day may be a strategy used by fungi to maximize fitness. 

However, releasing spores at the same time every day may not always be a good strategy; if turbulence is difficult to predict, rhythmic patterns of spore release may not maximize fitness. In our pool of simulations, the weakest diurnal cycle is found in location \#1 in Canada during the month of January (Figure 4A-D). Although January is not the typical season for sporulation, we use this simulation as an extreme example to illustrate maximum intermittency of turbulence. In this simulation, both turbulence and fitness vary irregularly from day to day and the indices $p_w$ and $p_\chi$ are close to zero (purple diamond in Figure 3D; Figure 4B,D). Every day is different. 
Even in environments where there is no periodicity, 
intense turbulence at liberation causes spores to be lifted up in altitude and fitness to plummet.  
This is exemplified in our most intermittent simulation, where negative fluctuations in fitness (troughs in Figure 4C) often occur when turbulence is intense (peaks in Figure 4A). 
To maximize fitness in intermittent environments, species may still evolve to liberate spores when turbulence is weak, but by evolving to measure turbulence directly.  

 To test this idea we compare two alternative models. In the first model, we consider fitness as a function of time of day (Figure 1 bottom row). In the second model, we consider fitness as a function of turbulence intensity (Figure 4E). In both cases, we perform non-linear regressions to obtain a function $f$ that predicts fitness $\chi=f(t)$ as a function of time of the day, and a second function $g$ that predicts fitness $\chi=g(w)$ as a function of turbulence (see Materials and Methods). We then compute the mean square error for the prediction in the two cases and for each of the 40 simulated locations and months, $\text{error(t)}=\sqrt{<[f(t) - \chi]^2>}$ and $\text{error(w)}=\sqrt{<[g(w) - \chi]^2>}$. We find that turbulence intensity predicts fitness more accurately than time of the day when the cycle is disrupted (purple region Figure 4F); time of the day and turbulence are nearly equally accurate in predicting fitness when the cycle is regular ($p_w\gtrsim0.8$, yellow region Figure 4F). In intermittent conditions, spores should be liberated whenever turbulence intensity is low regardless of whether it is day or night. This is especially true for our most intermittent simulation, which is an outlier (purple diamond in Figure 4F), but is also relevant to several other simulations where releasing spores according to turbulence may increase fitness by about 0.17 for example April in Colorado and July in Wisconsin. 
 
\begin{figure}[h!]
\begin{center}
\includegraphics[width=0.5\textwidth]{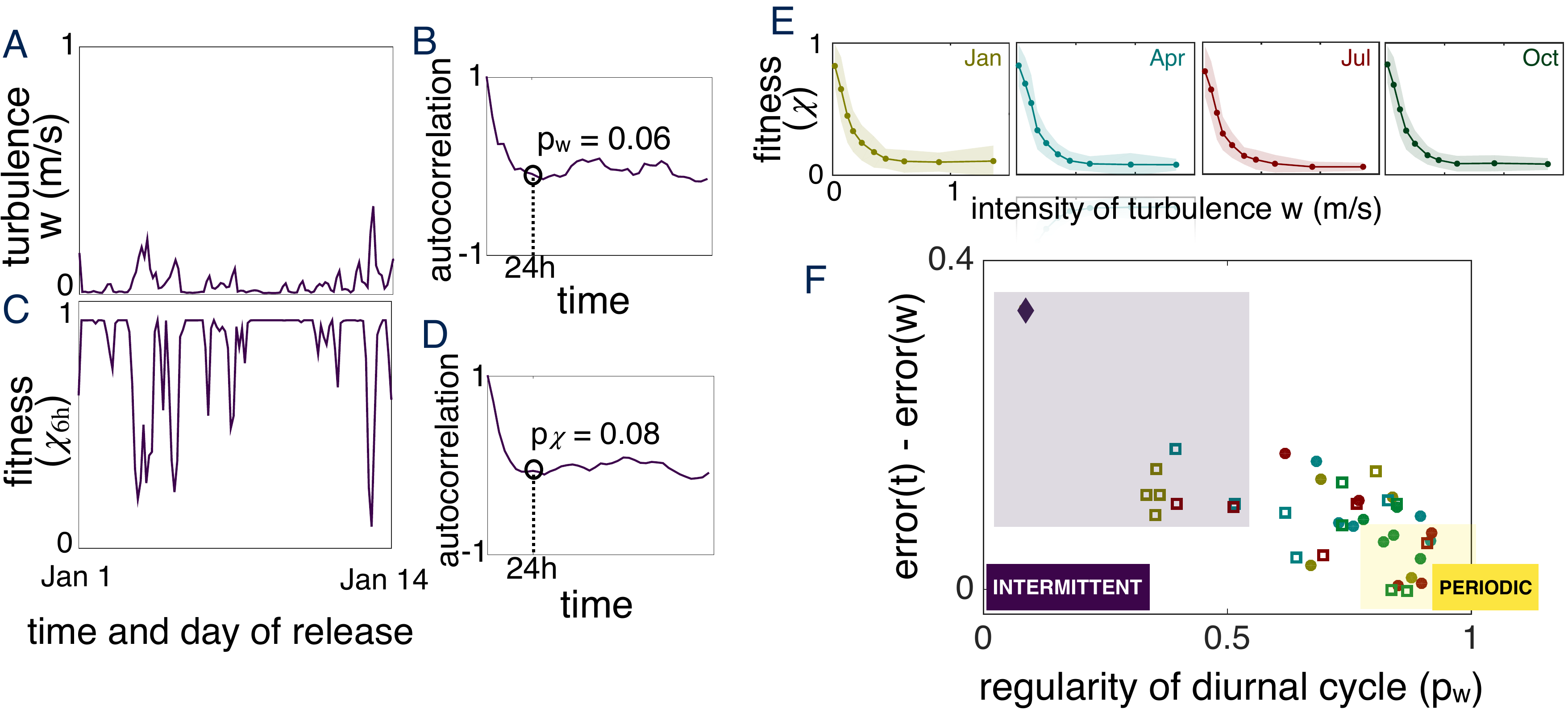}
\caption{
Releasing spores at the same time every day is not efficient when intensity of turbulence is intermittent. 
(A) Magnitude of vertical turbulent velocity $w$, (B) its autocorrelation function; (C) Fitness $\chi_{6h}$ and (D) its autocorrelation. Panels A-D are the same as Figure 3 A-B-D-E, for the same two weeks but a different location (\#1 in Canada, see Supplementary Table 1). 
(E) Fitness as a function of turbulence intensity, all simulations are pooled together by month. 
(F) Difference between error when predicting fitness according to turbulence \emph{vs} time of the day for each of the 40 simulated months. Squares correspond to the Northern locations (\#2 \#4 \#6 \#7 \#10) and dots the Southern Locations; months are color coded as in (E). The error is defined as $\sqrt{<[f(t)-\chi]^2>}$ for time of the day and $\sqrt{<[g(w)-\chi]^2>}$ for turbulence. 
Note how turbulence intensity is a better predictor of fitness than time of the day for small values of $p_w$ corresponding to intermittent conditions (purple shaded area). Whereas time of the day and turbulence are nearly equally accurate in predicting fitness when the atmosphere cycles regularly between weak and intense turbulence (yellow shaded region).
}
\end{center}
\end{figure}

\section{Discussion}
\noindent Our results demonstrate that time of the day when spores are released dramatically affects their fitness. Spores released during the day tend to be transported higher up in the atmosphere, and return to the ground after several days. But many spores can only survive few hours in the open atmosphere, and would die before returning to the ground. Hence, we predict that to maximize survival short-lived spores should be typically launched at night. \emph{Mycosphaerella fijiensis} \cite{Meredith_Lawrence1973,myco_book}, and \emph{Giberella zeae} \cite{Paulitz1996,Gilbert2008} release spores preferentially at night, and measures of their survival to UV light damage suggest that their lifetime in the open atmosphere is indeed extremely short, consistent with our prediction \cite{tau_myco1,tau_myco2,tau_gibbe1,tau_gibbe2}. \\ 
But why are some species releasing spores during the day? Our results show that fitness peaks at night for short-lived spores; however, 
fitness flattens out as spores are more long lived (see Supplementary Figure 3 for results with $\tau=2$ weeks and $\tau=2$ days). 
In other words, spores that are adapted to survive in the atmosphere for weeks can be released at any time of the day including when turbulence is maximum during the day.
Because survival in the air is not limiting, long lived spores may be released in conditions that maximize a different aspect of fitness, for example, distance travelled. In order to maximize distance travelled, these spores should be released in strong turbulent conditions, typically found during the day. Consistent with this hypothesis, asexual spores of \emph{Botyritis cinerea} \cite{Chastagner1978, Hartill1980}, \emph{Alternaria spp.}~\cite{OConnor2014}, \emph{Cladosporium spp.}~\cite{OConnor2014,Pady_Clary1969} preferentially release spores during daytime hours.\\

\noindent Releasing spores at specific time of the day modulates the expected fraction of spores surviving the journey, suggesting that fungi may tie spore liberation to an internal clock. Interestingly, it is known that spore production in \emph{Neurospora crassa} is indeed regulated by the circadian clock \cite{circadian,circadian2}, but whether this may provide any selective advantage has been hitherto unclear. Our results suggest that producing spores at certain times of the day may be instrumental to release them at times that maximize chances of survival. \\
 
\noindent A clock is not always a useful predictor of fitness: in cases where the atmospheric cycle is disrupted, turbulence varies irregularly, and releasing spores at the same time every day can lead to massive losses. 
In regions where the cycle is disrupted,  short lived spores should be released intermittently, whenever turbulence is weak, whereas long lived spores can be released in conditions of intense turbulence. 
Intermittent patterns of spore liberation do occur \cite{aylor_annurev}, however what features of the environment dictate spore release is still poorly understood \cite{review_effects_meteo}.
The abscission of asexual spores of, e.g., \emph{Cochliobulus heterostrophus} (syn. = \emph{Helminthosporium maydis}), whose spores are among those considered to be capable of withstanding atmospheric conditions experienced during continental-scale dispersal \cite{Mundt2009}, occurs only when wind velocity exceeds 10 m/s \cite{Aylor1975}. This is consistent with our hypothesis, because large air speed close to a substrate is usually associated to intense turbulent wind gusts. \\
 
\noindent 
Whether a clock-based or a sensation-based strategy is more effective, depends on the environment that a species is adapted to. 
To compare these two alternative strategies, we introduced the parameter $p_w$ which distinguishes between regions with regular  \emph{vs} intermittent atmospheric conditions. 
This analysis provides a testable hypothesis for the emergence of intermittent \emph{vs} regular spore release. If patterns of spore liberation are shaped by the need to maximize spore survival in the atmosphere, species adapted to regions where the diurnal cycle is strong ($p_w>0.8$ in our test case) will release spores according to their internal clock. Whereas species adapted to an environment with a weak diurnal cycle ($p_w<0.8$ in our test case) are more likely to liberate spores intermittently in response to environmental conditions that measure turbulence intensity. 
The value of $p_w$ that marks this transition will vary from species to species, according to spore morphology and longevity, but the qualitative pattern is robust. 
To test this hypothesis, the patterns of spore liberation must be monitored in the field, along with the genetic structure of the population and spore morphology and longevity in the atmosphere. What species favors one strategy over the other will determine how spore liberation will shift in response to environmental changes including global change.\\

\noindent 
Spore dispersal is generally regarded as dangerous and fundamentally wasteful. Our results demonstrate that although fungi do lose control over individual spores, they can still maximize fitness in the atmosphere by manipulating the timing for spore release.
Indeed, the timing for spore liberation dictates the fraction of spores that survive their journey in the open atmosphere. In other words, the ensemble statistics of spore flight time keeps memory of the initial conditions that spores meet when they first reach the open air. 
These results partially reconcile two contrasting aspects of fungal spore dispersal: microscopic optimization \emph{vs} large scale uncertainty. On the one hand, at micron to cm scales, fungi evolved fascinating adaptations to maximize efficiency of the microscopic mechanism of discharge (reviewed in \cite{annurev}). The ascomycetes, the most numerous phylum in the higher fungi, fire sexual spores from a pressurized cell which is finely regulated to minimize dissipation\cite{marcus_shape,apical_ring,puffing,review_frances,book_chapter}. The basidiomycetes, including $\sim 40000$ mushroom-forming species, eject spores through a surface tension catapult that achieves precise control of spore range right after discharge~\cite{Buller1909,Ingold1939,Turner&Webster91,Pringle2005b,Noblin2009,Stolze-Rybczynski2009,liu2017}. These adaptations indicate that spore dispersal is under considerable selective pressure. But this idea is at odds with the fact that, once spores reach dispersive airflows, their fate is dictated by a series of stochastic events and appears entirely out of control. Fungi acknowledge uncertainty by producing a large number of propagules, making fungal migration extremely different from the ordered migration of mammals and fundamentally wasteful \cite{ordered_migration}. Here we show that, in addition, fungi may use their exquisite microscopic control to release spores when their chances of survival in the atmosphere are highest.\\
    
\section{Lagrangian simulations with meteorological data using HYSPLIT}
To compute the statistics of flight times, we follow many particles released from a given location at different times using the Hybrid Single Particle Lagrangian Integrated Trajectory (HySPLIT) model \cite{ref:Hysplit} an open source code developed at the Air Resource Laboratory of the National Ocean and Atmospheric Administration (ARL-NOAA) in the US. We modeled spores as passive tracers with an additional gravitational settling velocity. Spores are transported by the wind, whose velocity is obtained from meteorological datasets on a large scale grid with resolution 32~$km$, turbulence on smaller scales is modeled as a correlated stochastic process, so spore trajectories can be computed through a Langevin equation:
\begin{equation} \label{eq:langevin2}
\frac{d \mathbf P(t)}{dt} = \mathbf  V_{\text{meteo}} [ \mathbf P (t),t] +  \mathbf  V'[ \mathbf P (t),t]  + \mathbf V_G.
\end{equation} 
\noindent where $\mathbf P$ is the three dimensional instantaneous location of a spore, $ \mathbf V_{\text{meteo}}$ is the large-scale wind velocity from the North American Regional Reanalyses (NARR)\cite{ref:Narr}; $ \mathbf  V'$ is a realization of the stochastic turbulent fluctuation \cite{ref:Wilson} and $\mathbf V_G$  the gravitational settling velocity. 
NARR is an extended dataset of meteorological variables on a regular grid covering the whole North American continent resulting from a matching procedure between outputs of numerical models and sparse observations of many atmospheric variables. NARR data are given on a Lambert conformal grid with 309 x 237 horizontal points and 24 levels on a vertical pressure-sigma coordinates system. The nominal horizontal resolution is $32$ km. Starting from 1979 to today the state of the atmosphere is available at a time resolution of $3$ hours. 
The variance of turbulent fluctuations in the vertical direction depend on height and is modeled with semi-empirical expressions that vary with the state of the atmosphere, i.e.~stable \emph{vs} unstable (see Supplementary Information for details, \cite{ref:KCclo,ref:BHclo}). The specific choice for the closure as well as their dependence on altitude only weakly affect fitness (see Supplementary information). 

\noindent Spores sediment with a constant downward velocity. Qualitative results are robust for sedimentation velocities of the order of $1\,cm/s$. As a reference, this value of sedimentation velocity corresponds to a sphere with the density of water and diameter $18\,\mu m$. Results vary considerably for spores sedimenting $10$ times faster (see Supplementary Information). 
Finally, dry deposition to the ground is computed assuming that the flux of spores $j(\mathbf{x},t)$ to the ground is proportional to concentration of spores close to the soil $\theta( \mathbf x,t)$: $j(\mathbf x,t)=V_d\theta( \mathbf x,t)|_{z=0}$,
where $V_d$ was taken equal to the gravitational settling, as appropriate for spherical particles larger than about $1\,\mu m$ \cite{aerosol_book}. A more detailed modeling of deposition on the canopy including dependence on spore shape is left for future studies.
Figure 2 shows that deposition velocities affect fitness quantitatively, but do not affect the qualitative patterns. 
In our model, the gravitational/deposition velocity is the only parameter of the dynamics that depends on the fungal species. 
Wet deposition may decrease the flight time in daytime releases, but does not change the general conclusions unless fungi are able to release spores right before rain. There is some evidence that this may be true for some species, and we will treat this fascinating possibility elsewhere. 

\section{One dimensional model of spore transport}
In the Eulerian framework, the concentration of passive tracers advected by a short-correlated velocity field in one dimension follows the well-known Fokker--Plank equation $\partial_t\theta(z,t) = \partial_z [D(z)\partial_z\theta]+v_D\partial_z\theta $, see e.g.~\cite{okubo,fgv}, where $\theta(z,t)$ is the average concentration of spores at altitude $z$ and time $t$; $D(z)$ is the vertical eddy diffusivity, for which we use the closures implemented in the HYSPLIT simulations (see Supplementary information), and $v_D$ is the sedimentation velocity. We compared fitness obtained through this simplified one-dimensional model that neglects the horizontal dynamics entirely to the results of the full HYSPLIT model. The one dimensional model captures well the importance of stability in determining fitness, and reproduces the oscillations observed in the full simulations. We impose reflecting boundary conditions at the top of the domain (25 km), and we remove a fraction of spores localized at $z< 50\, \mathrm{m}$ according with the deposition velocity, as done by the HYPLIT model. We adopt a finite volume scheme with regular cells of $\Delta z = 5\, \mathrm{m}$ and a Runge-Kutta algorithm of fourth order for time marching, with time step $\Delta t$ chosen according to the Courant criterion: $\Delta t = \min_{D_i} {\frac{\Delta z^2}{8D_i}}$ where $D_i = D(z_i)$ is the eddy diffusivity evaluated at the center of the i-th vertical cell.

\section{Regression}
To determine whether time of the day or vertical turbulence hold the most reliable information about fitness, we perform non linear regression using Gaussian processes \cite{mccay, bishop, rasmussen_williams}. This is a non-parametric method for regression, with the advantage that the fitting function can be represented as an infinite sum of basic functions hence making the procedure extremely flexible. We target a function that predicts fitness $\chi$ from input data $x$, where $x$ represents either time of day or turbulent intensity at spore release. $\chi$ is assumed to be a Gaussian random variable, with mean $\bar{\chi} = f(x)$ and variance $\beta^{-1}$. 
A kernel function is defined to quantify distances between different input points, here we use a Gaussian kernel $k(x_i,x_j) = e^{-||x_i-x_j||^2/(2\pi\lambda^2)}$. 
Given a training set $(\mathbf x,\bm{\chi})=(x_1,..,x_N,\chi_1,...,\chi_N)$, the probability distribution $p(\chi_{N+1}|\bm{\chi})$ for the value of $\chi$ associated to a test data $x_{N+1}$, conditioned to the previous observations, is a gaussian with mean 
$$f(x_{N+1})=\mathbf k^T \underline{\underline{C}}^{-1} \bm \chi$$
and variance 
$$\sigma^2(x_{N+1}) = c -\mathbf k^T \underline{\underline{C}}^{-1} \mathbf k$$
\noindent where $C(x_n, x_m) = k(x_n, x_m) + \beta^{-1}\delta_{nm}$; $c = k(x_{N+1}, x_{N+1}) +\beta^{-1}$ and $\mathbf k =k(x_{N+1},x_n)$ and $n,m \in (1,N)$.
We choose the parameters of the model, $\lambda$ and $\beta$, so as to minimize the generalization error using cross validation. We split our 9600 simulations (10 locations $\times$ 4 months $\times$ 30 days $\times$ 8 releases per day) in 240 test points (1 location $\times$ 1 month $\times$ 30 days $\times$ 8 releases per day) and  9630 training points (the rest). We use a random subsample of 1000 (for $t$) and 1500 (for $w$) simulations from the training set to train the algorithm and use it to predict the test set. We  compute the generalization error on the test data and we repeat the procedure over the 40 different ways to split the dataset in test and training, and average to obtain the test error. We repeat the procedure varying systematically the parameters and identify the region in the parameter space that provide minimum test error. The difference in the generalization error, defined as $(\chi-f(x))^2$, for $x = $ time of day and $x = $ turbulence intensity, with the parameters selected as above, is plotted in Figure 4F.


\begin{acknowledgments}
This work was supported by the Agence Nationale de la Recherche Investissements d'Avenir UCA$^{\textrm{\sc \sf \tiny JEDI}}$ \#ANR-15-IDEX-01 and CNRS PICS ``2FORECAST'', by the Thomas Jefferson Fund, a program of FACE. We acknowledge support for computational resources from INFN and CINECA. The authors gratefully acknowledge the NOAA Air Resources Laboratory (ARL) for the provision of the HYSPLIT transport and dispersion model. NCEP Reanalysis data were provided by the NOAA/OAR/ESRL PSD, Boulder, Colorado, USA, from their Web site at 
https://www.esrl.noaa.gov/psd/
\end{acknowledgments}

\end{article}

\end{document}


\title{Supplementary Information for: ``Timing of fungal spore release dictates survival to atmospheric transport''}

\author{Daniele Lagomarsino Oneto}
\affiliation{Universit\'e C\^ote d'Azur, Institut de Physique de Nice, 06108, Nice France}
\author{Jacob Golan}
\affiliation{Departments of Botany and Bacteriology, University of Wisconsin-Madison, Madison, WI, USA}
\author{Anne Pringle}
\affiliation{Departments of Botany and Bacteriology, University of Wisconsin-Madison, Madison, WI, USA}
\author{Andrea Mazzino}
\affiliation{Department of Civil, Chemical, and Environmental Engineering, University of Genova and INFN, via Montallegro 1, 16145 Genova, Italy}
\author{Agnese Seminara}
\affiliation{CNRS and Universit\'e C\^ote d'Azur, Institut de Physique de Nice, Parc Valrose, 06108, Nice} 

\begin{abstract} This document includes the following Supplementary Information for the paper: ``Timing of fungal spore release dictates survival to atmospheric transport''
\begin{itemize}
\item[\ref{sec:hysplit}] Lagrangian simulations with weather data using HYSPLIT

\end{itemize}
\end{abstract} 

\maketitle


\begin{figure}[h!]
\includegraphics[width=0.75\textwidth]{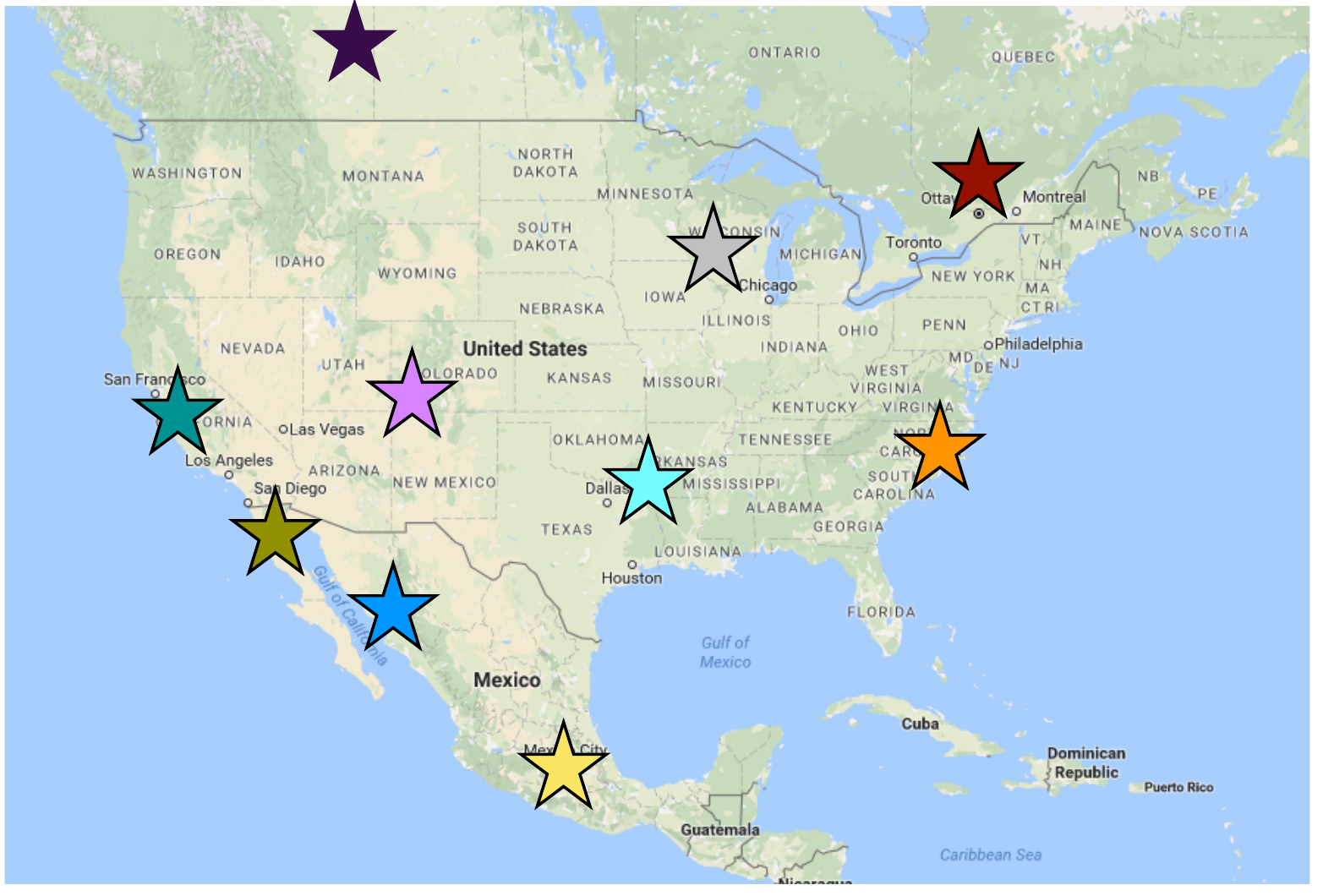}
\caption{\footnotesize Map of starting locations for our simulations.}
\label{fig:map}
\end{figure}
\begin{figure}[h!]
\includegraphics[width=\textwidth]{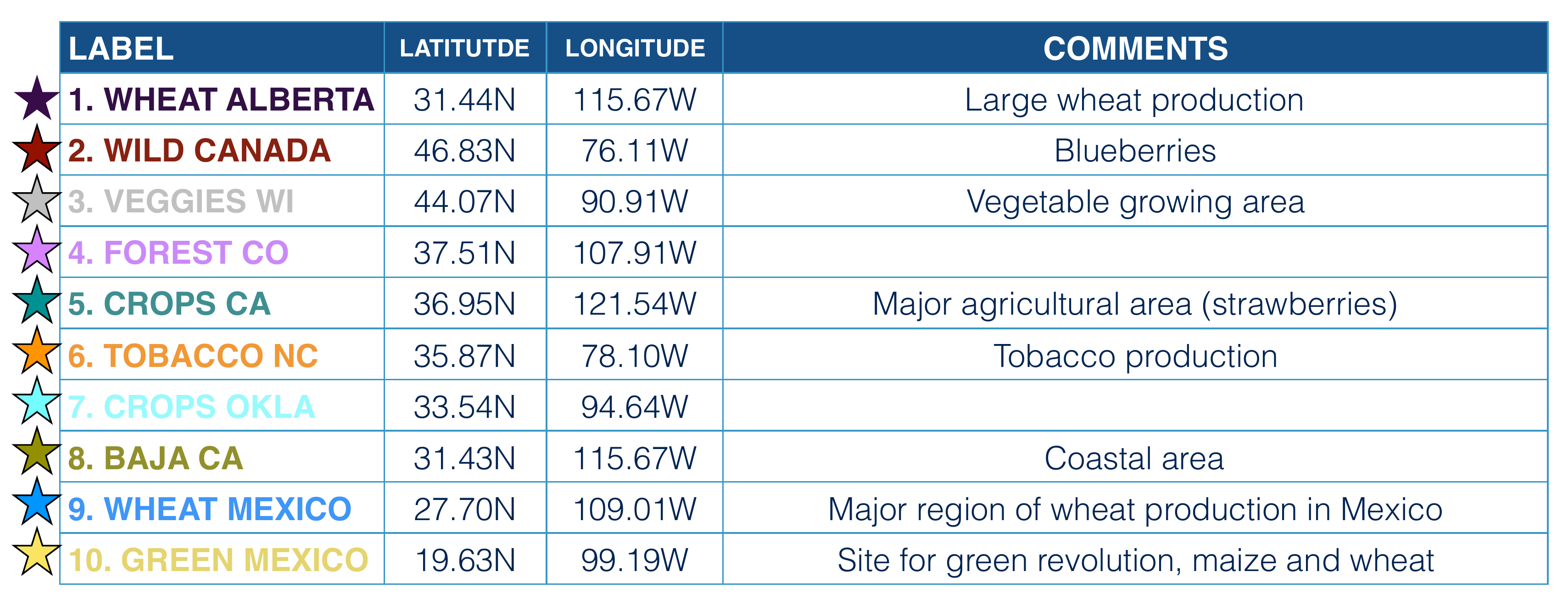}
\caption{\footnotesize Details of starting locations.}
\label{fig:table}
\end{figure}
\begin{figure}[h!]
\includegraphics[width=\textwidth]{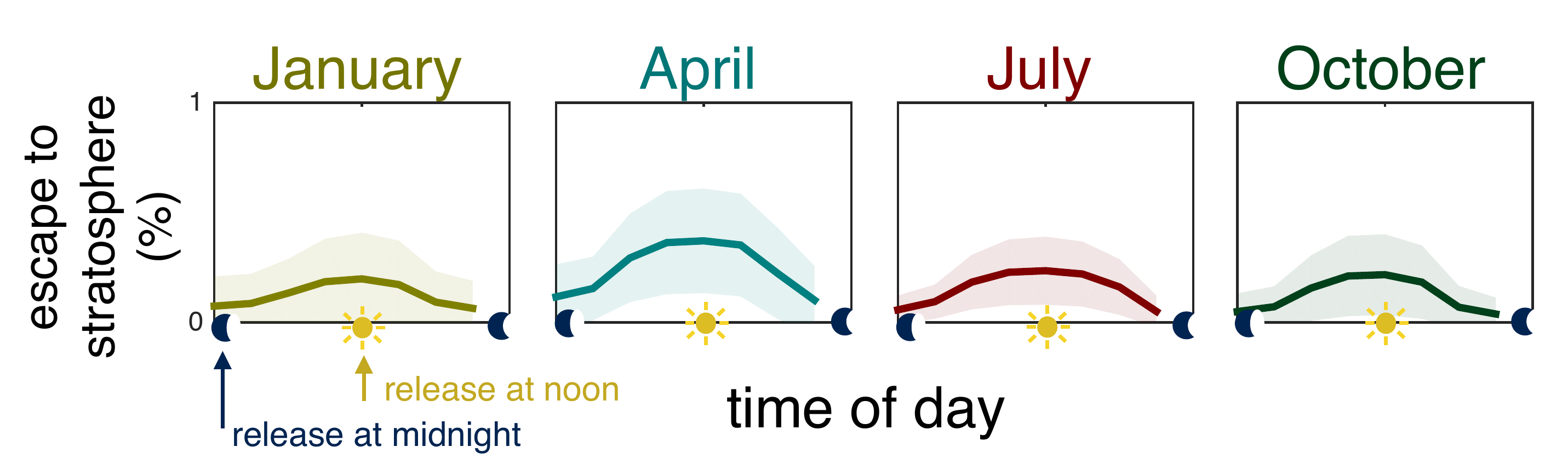}
\caption{\footnotesize A small fraction of spores escapes into the stratosphere, especially during highly convective conditions typically found during the day. Left to right: average fraction of spores escaping into the stratosphere as a function of time of the day for the 4 months we simulated. Averaged over the 10  starting locations.}
\label{fig:strato}
\end{figure}
\begin{figure}[h!]
\includegraphics[width=\textwidth]{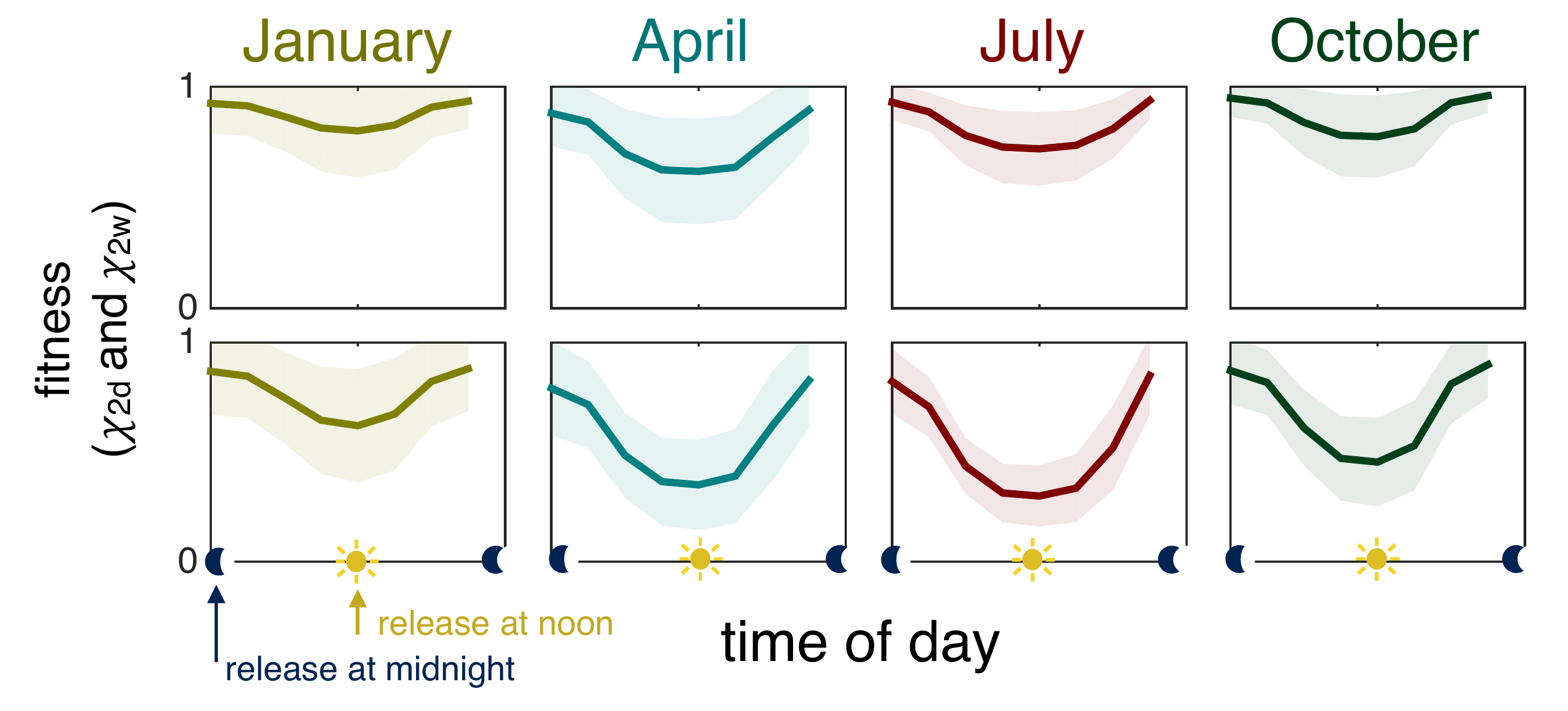}
\caption{\footnotesize Fitness for longer lived spores follows similar patterns as those described in the main text. Selective pressure is released for longer lived spores that suvive flights in most conditions. Top row: fitness for spores with lifetime 2 weeks, Bottom row: fitness for spores with lifetime 2 days.}
\label{fig:tau}
\end{figure}
\begin{figure}[h!]
\includegraphics[width=\textwidth]{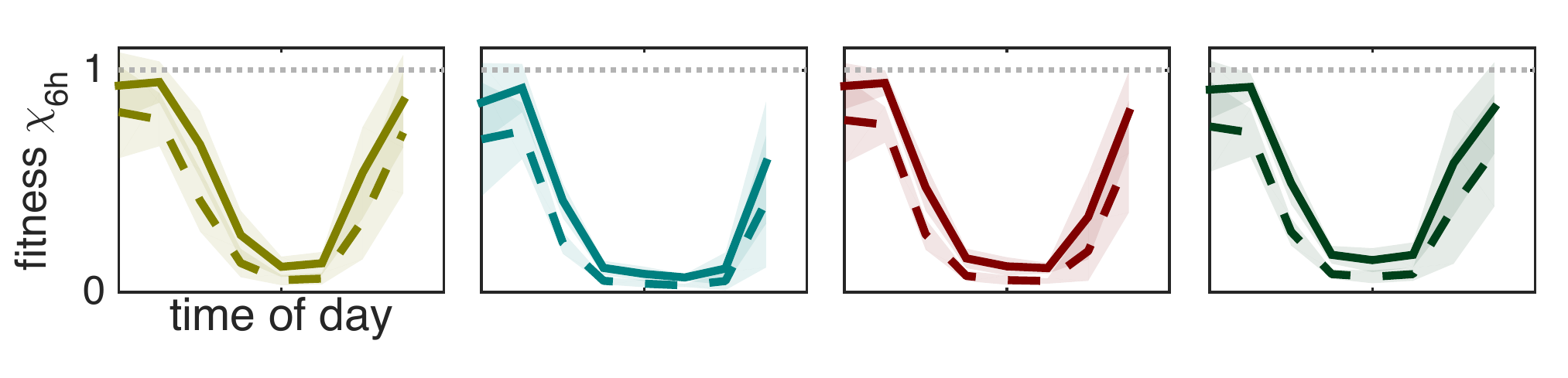}
\caption{\footnotesize Results vary with settling velocity. From left to right: fitness for spores with lifetime $6$ hours for the 4 months simulated as described in the main text. Full simulation described in the main text with settling velocity $6 \, mm/s$ (colored dashed lines); with double settling velocity $12\, mm/s$ (colored solid lines); settling velocities of $6\,cm/s$, 10 times larger than the simulations showed in the main text, causes spores to always sediment in less than 6 hours (dotted gray lines). Patterns in fitness occur even for these large spores, if a shorter lifetime is considered (Data not shown). }
\label{fig:vd}
\end{figure}
\begin{figure}[h!]
\includegraphics[width=\textwidth]{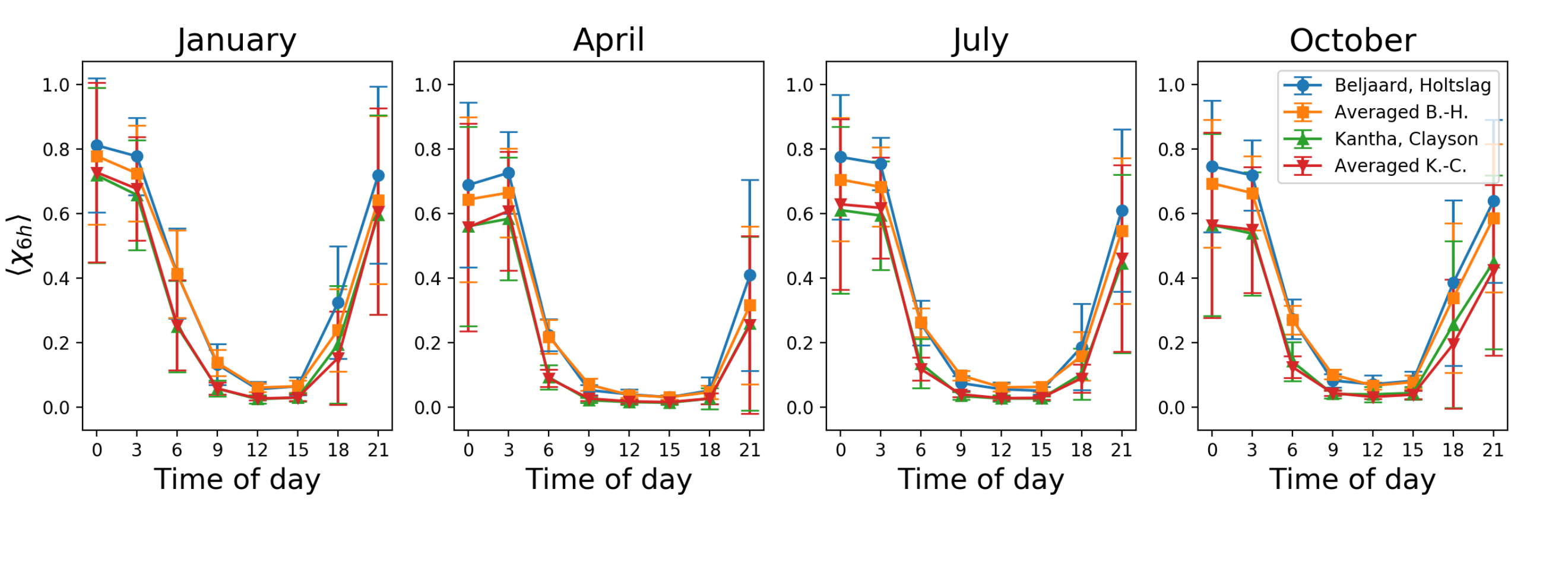}
\caption{\footnotesize Robustness of results with respect to different closures for modelisation of turbulent fluctuations of vertical wind. Left to right: Fitness of spores with lifetime $6\,h$, for the months of January, April, July and October. Different colors / symbols correspond to different closure schemes:  Beljaars and Holtslag with the $z$-dependence (blue dots); Beljaars and Holtslag averaged over $z$ (green squares); Khantar and Clayson with the $z$-dependence (green upward triangles); Khantar and Clayson averaged over $z$ (red downward triangle).}
\label{fig:closures}
\end{figure}

\clearpage
\newpage

\section{Lagrangian simulations with weather data using HYSPLIT}
\label{sec:hysplit}
In this work we simulate atmospheric transport using the Lagrangian approach, i.e.~we follow the trajectories of many individual particles rather than describing the evolution of the concentration of particles.
We model spores as passive tracers with an additional gravitational settling velocity, which is a commonly used approximation in geophysical transport. Neglecting inertial effects is justified for typical spores (about $10\,\mu m$ in radius and density of water) whose Stokes timescale is about $1$~ms, much smaller than the Kolmogorov timescale, the smallest characteristic timescale for turbulent transport, which is $\tau_\eta \sim 0.1$~s. 
Stokes numbers of the order of 0.01 result in cancellation of many terms in the general Maxey-Rayley form of the equation of motion for spheres immersed in a fluid\cite{ref:Maxey}. Under these assumptions spore trajectories are solutions of the following ordinary differential equation:
\begin{equation} \label{eq:langevin}
\frac{d \mathbf P(t)}{dt} = \mathbf  V_{\text{wind}} [ \mathbf P (t),t] + \mathbf V_G,
\end{equation}
\noindent where $\mathbf{P}(t)$ is the position of the particle at time $t$; $\mathbf{V}_{\text{wind}}[ \mathbf P (t),t]$ is the instantaneous velocity of the wind at the location occupied by the particle at time $t$ and $\mathbf V_G$ is the gravitational settling. 
A complete description including all dynamical degrees of freedom of the wind velocity field from the smallest to the largest scales ($mm$ to hundreds of $km$) is practically inaccessible. To overcome the problem, HYSPLIT takes a widely used approach, and describes the flow from the largest scale down to a certain spatio-temporal resolution, whereas the structure of the velocity field below this intermediate scale is modelled. The large-scale velocity of the wind is provided by regional meteorological models and a stochastic term describes the small-scale turbulence which is in general neither stationary nor homogeneous. Spore trajectories can be computed through a Langevin equation:
\begin{equation} \label{eq:langevin2}
\frac{d \mathbf P(t)}{dt} = \mathbf  V_{\text{meteo}} [ \mathbf P (t),t] +  \mathbf  V'[ \mathbf P (t),t]  + \mathbf V_G.
\end{equation} 
Here $ \mathbf V_{\text{meteo}}$, $ \mathbf  V'$ and $\mathbf V_G$ are the large-scale wind velocity from meteorological datasets; the turbulent velocity fluctuations and the gravitational settling velocity respectively. 
Lagrangian simulations described in the maintext are performed using the Hybrid Single Particle Lagrangian Integrated Trajectory (HySPLIT) model \cite{ref:Hysplit}, an open source code developed at the Air Resource Laboratory of the National Ocean and Atmospheric Administration (ARL-NOAA) in the US.
\paragraph{Numerical technique} To find a numerical solution of Eq.~\ref{eq:langevin2} we consider the following equivalent discrete-time set of equations:
$$
\begin{aligned}
&X(t+\Delta t) = X_\text{Adv}(t+\Delta t) + U'(t+\Delta t)\Delta t\, G_X  \\
&Y(t+\Delta t) = Y_\text{Adv}(t+\Delta t) + V'(t+\Delta t)\Delta t \,G_Y  \\
&Z(t+\Delta t) = Z_\text{Adv}(t+\Delta t) + W'(t+\Delta t)\Delta t\, G_Z  + V_G\Delta t 
\end{aligned}
$$
where $\Delta t$ is the time step; $(X,Y,Z)=\mathbf P$ are the three components of spore position; the pedices $\text{Adv}$ indicates new positions computed after advection by the large-scale wind speed, $(U', V', W')=\mathbf  V' $ are the turbulent velocity components and $G_X, G_Y, G_Z $ are scaling factors depending on the adopted coordinates. 
Wind fields for advection are taken from the North American Regional Reanalyses (NARR)\cite{ref:Narr}, which is an extended dataset of meteorological variables on a regular grid covering the whole North American continent. This dataset is the result of a matching procedure between outputs of numerical models and sparse observations of many atmospheric variables. NARR data are given on a Lambert conformal grid with 309 x 237 horizontal points and 24 levels on a vertical pressure-sigma coordinates system. The nominal horizontal resolution is $32$ km. Starting from 1979 to today the state of the atmosphere is available at a time resolution of $3$ hours. 
The same system of horizontal coordinates as that of the  meteorological dataset is used; fields are interpolated on an internal sigma-terrain-following system in the vertical direction, whose coordinates follow the orography of the terrain.
The time marching for advection is made in two steps, initially computing a first guess position:
$$
\mathbf P'(t+\Delta t) = \mathbf P(t) + \mathbf V(\mathbf P,t)\Delta t,
$$
where $\mathbf V(\mathbf P,t)$ is the resolved wind velocity at time $t$ interpolated at the particle position and then giving a final position
$$
\mathbf P(t+\Delta t) = \mathbf P(t) + \frac{1}{2}\left[ \mathbf V(\mathbf P,t) + \mathbf V(\mathbf P',t+\Delta t)\right]\Delta t.
$$
As mentioned before, turbulent velocities are realizations of stochastic processes \cite{ref:Wilson}. For each component such processes are described by the following equations:
$$
\begin{aligned}
&U'(t+\Delta t) = R(\Delta t) U' (t) + U'' \sqrt{1-R(\Delta t)^2}\\
&V'(t+\Delta t) = R(\Delta t) V' (t) + V'' \sqrt{1-R(\Delta t)^2}\\
&\left(\frac{W'}{\sigma_W}\right)(t+\Delta t) = R(\Delta t) \left(\frac{W'}{\sigma_W}\right)(t) +\\
&+ \left(\frac{W''}{\sigma_W}\right)(t)  \sqrt{1-R(\Delta t)^2} + T_{L_W}(1-R(\Delta t))\frac{\partial \sigma_3(t)}{\partial z}
\end{aligned}
$$
with 
$$
\sigma_W(t+\Delta t) = \sigma_W(t) +W'(t) \Delta t  \frac{\partial \sigma_3(t)}{\partial z}
$$
and where the autocorrelation function $R$ is
$$
R(\Delta t) =  e^{-\frac{\Delta t}{T_{L_i}}}
$$ 
with $T_{L_i} = T_{L_W}$, $T_{L_U}$ or $T_{L_V}$ Lagrangian time scales, assumed as constant in the model and equal to $T_{L_W}  = 100 s $ and $T_{L_U}  = T_{L_V} =10800 s $. 
The random velocities $U'',\, V'',\, W''$ are computed as 
$$
U'' = \sigma_1 \lambda_1,\quad V'' = \sigma_2 \lambda_2. \quad W'' = \sigma_3 \lambda_3,
$$
where $\lambda_i$ are realizations of gaussian random processes with mean $0$ and unitary standard deviation and the $\sigma_i^2$ are the turbulent velocity variances which values are function of both time and space and depend on the turbulent parameterization (turbulence closure) adopted, discussed below. 

For the horizontal components of the velocity the problem is closed connecting turbulent velocity variances with the deformation gradient of the large scale wind field:
$$
\sigma_{1,2} =\sqrt{ \frac{1}{\sqrt{2}T_{L_U}}( c \Delta X)^2 \left[ \left( \frac{\partial v }{\partial x} +  \frac{\partial u }{\partial y} \right) ^2 + \left(  \frac{\partial u }{\partial x} - \frac{\partial v }{\partial y} \right)^2 \right] ^ {\frac{1}{2}}}
$$
where $c= 0.14$ and $\Delta X$ is the horizontal grid step \cite{smagorinski,deardoff}

Vertical turbulence requires a preliminary evaluation of the atmospheric boundary layer (ABL) stability, which is done by HySPLIT by evaluating the \textit{Monin-Obukhov length} $L$, that is a dimensional parameter related with the stability of the surface layer, the lower part of the ABL. It is positive or negative when the ABL state is respectively stable or unstable. The expression for $L$ is
$$
L = -\frac{(u^\ast)^3}{kg \frac{\langle w'\theta' \rangle}{\theta}}.
$$
The parameter $u^\ast = [\langle u'w'\rangle^2 + \langle v'w'\rangle^2]^{\frac{1}{4}} = |\frac{\tau_s}{\rho}|^\frac{1}{2}$, called the \textit{friction velocity}, is related with the turbulent stress at the surface $\tau_s$ and the air density $\rho$; the covariances $\langle u'w'\rangle$ and $\langle u'w'\rangle$ are the kinematic momentum fluxes at the surface. At the denominator $g$ is the gravity acceleration, $k=0.4$ is the von Karman constant, $\theta$ is the potential temperature and $\langle w'\theta' \rangle$ is proportional to the turbulent heat flux at the surface. From a physical point of view, $L$ measures the relative contributions to turbulent kinetic energy given by buoyant production and shear production, representing the heigth at which these two contributions are equal. \\ 
During unstable conditions another velocity scale can be introduced, the convective velocity $W^\ast = \left[ \frac{g Z_i}{\theta}\langle w'\theta' \rangle\right]^\frac{1}{3}$, where $Z_i$ is the boundary layer depth, that is another important parameter involved in the computation of turbulent motions.
All these parameters are internally evaluated by the model from meteorological data (for more details see \cite{ref:Hysplit}) and they are involved in the computation of vertical turbulent velocities which are obtained by semi-empirical formulas. Several possible semi-empirical parameterizations, also called closures, can be chosen. 
A possible choice is to follow Khantar and Clayson equations \cite{ref:KCclo}, according to which the vertical velocity variances during stable conditions are given by
$$
\begin{aligned}
&\sigma_3^2 = 3.0 (u^\ast)^2 &(\mathrm{surface\, layer})\\
&\sigma_3^2 = 3.0 (u^\ast)^2\left(1-\frac{z}{Z_i}\right)^{\frac{3}{2}} &\mathrm{(rest\, of \,the \,ABL)},\\
\end{aligned}
$$
while unstable conditions are modeled as follows:
$$
\begin{aligned}
\sigma_3^2 = 1.74 (u^\ast)^2   \left(1-3\frac{z}{L}\right)^{\frac{2}{3}}&(\mathrm{surface\,layer}) \\
\sigma_3^2 = 3.0 (W^\ast)^2  \left(\frac{z}{Z_i}\right)^{\frac{2}{3}} \left(1-\frac{z}{Z_i}\right)^{\frac{2}{3}}  \times
\left(1+0.5 R^{\frac{2}{3}}\right) &\mathrm{(rest\, of \,the \,ABL)}\\
\end{aligned}
$$
where $R = 0.2$. \\
An alternative approach is given by Beljaars and Holtslag \cite{ref:BHclo}, which identifies the diffusion coefficient for particles with the diffusion coefficient for heat, and then converts them into turbulent velocities. According to this idea the vertical diffusion coefficient $K_3$ is given by the following expression:
$$
K_3 = kw_hz \left(1-\frac{z}{Z_i}\right)^2,
$$
where the stability parameter $w_h$ changes according to the stability of the ABL and is a function of friction velocity, Monin-Obukhov length, and convective velocity as described in \cite{manual_Hysplit}.
Therefore, once diffusion coefficients have been computed, the turbulent velocity variances result from
$$
\sigma_3 = \sqrt{\frac{K_3}{T_{L_W}}}.
$$
Both parameterizations provide vertical turbulent velocities as a function of the altitude. We consider the further possibility to first compute the vertical profiles of turbulent diffusivities at each $x,y$ location, and then averaging them over the ABL, so that the vertical fluctuation does not depend on height above the ground (a condition that we have labeled ``vertically uniform'' in Figure 2 of the main paper).  \\
Analogous semiempirical descriptions are used also for the stratosphere, above the ABL, where turbulent velocities quickly become negligible and their contribution is subdominant in the dynamics with respects to the stronger large scale circulation. 
The main results showed in the main text are obtained using the Beljaars and Holtslag scheme.
We have tested the robustness of our results by changing the parameterization of vertical turbulence, and we find that typical patterns of our statistical observables are preserved in all cases (see Figure~\ref{fig:closures}).

A further player in in the dynamics is gravitational sedimentation, described by a constant downward velocity. Approximating spores as spheres with the same volume of the spore, the sedimentation velocity can be computed starting from spore equivalent radius $r$ and its density $\rho_S$ 
$$
V_{G}= g\tau_S = \frac{2}{9} \frac{(\rho_S-\rho_{Air}) gr^2}{\mu}
$$
where $\tau_S$ is the Stokes time, $\mu = 1.81\, 10^{-5} \,Pa\cdot s $ is the dynamical viscosity of air, $\rho_S$ and $\rho_\text{Air}$ are spore and air densities respectively. 
This is the only parameter of the dynamics that depends on the fungal species, specifically on spore size, shape and density. 
We set this parameter to $V_G=6\,mm/s$, resulting from characteristic parameters of the wheat pathogen \textit{Puccinia Graminis} that is largely present in the North American continent. Althought this is a specific choice, many species have similar sedimentation velocities. Moreover, gravitational settling affects fitness significantly less than vertical turbulence as shown for $\chi_{6h}$ in Figure 2 of the main text, and Figure~\ref{fig:vd}.

The last factor affecting residence times is dry deposition to ground which is computed assuming that the flux $j(\mathbf x,t)$ of spores to the ground is proportional to concentration $\theta( \mathbf x,t)$ close to the soil:
$$j(\mathbf x,t)=V_d\theta( \mathbf x,t)|_{z=0}$$
where the parameter $V_d$ has the dimension of a velocity and has been taken equal to the gravitational settling velocity, as appropriate for particles larger than about $1\,\mu m$ \cite{deposition}.
Assuming that deposition occurs in a layer of thickness $\Delta Z_d$ above the ground level and taking particle concentration as constant within this layer, the time employed by the mass present in $\Delta Z_d$ to be completely deposited is equal to $\frac{\Delta Z_d}{V_d}$ and hence the fraction of mass deposited in a time step $\Delta t$ is given by the ratio $\frac{V_d\Delta t}{\Delta Z_d}$.
This mass drop is reproduced removing randomly particles below $\Delta Z_d$ with probability $$P = \frac{V_d\Delta t}{\Delta Z_d}$$.
As pointed out above and showed in Figure 2 of the main text, deposition velocities affect fitness quantitatively, but do not affect the qualitative patterns observed. 
Wet deposition will affect the results, but only modifies the picture if fungi are able to release spores when rain events are more probable. Although this is a fascinating possibility, and some evidence is starting to emerge in this direction, this remains outside the scope of the present work and will be examined in detail elsewhere.

\begin{table}[ht]\centering\caption{Comparison of physical spore parameters for two species.}
\label{tab:spores}
\begin{tabular}{|cccc|}
\hline
Species &size $[\mu m]$& density $[\frac{g}{cc}]$ & Stokes time $[s]$\\ 
\hline \hline
\textit{Puccinia graminis}& $28.3$ (length) & $0.47$ & $1.78 \, 10^{-3} $  \\
	& $17.5$ (width) &   &    \\
\textit{sclerotinia sclerotiorum}& $12$ (length) & $0.44$ & $1.95 \, 10^{-4} $  \\
	& $6$ (width) &   &    \\
 \hline
 \end{tabular}
\end{table}